\documentclass[runningheads]{llncs}
\usepackage{graphicx}
\usepackage{comment}
\usepackage{amsmath,amssymb} % define this before the line numbering.
\usepackage{color}
\usepackage{cite}
\usepackage{multirow}
\usepackage{subfigure}
\usepackage{epstopdf}
\usepackage{caption}
\usepackage{booktabs}
\usepackage{float}
\usepackage{amsmath,bm}
\usepackage{mathrsfs}
\usepackage{url}
\usepackage{enumitem}
\DeclareMathOperator*{\argmin}{arg\,min}

\begin{document}
% \linenumbers
\pagestyle{headings}
\mainmatter

\title{Deep Spatial-angular Regularization for Compressive Light Field Reconstruction over Coded Apertures} % Replace with your title

%\begin{comment}
% \titlerunning{ECCV-20 submission ID \ECCVSubNumber} 
% \authorrunning{ECCV-20 submission ID \ECCVSubNumber} 
% \author{Anonymous ECCV submission}
% \institute{Paper ID \ECCVSubNumber}
%\end{comment}
%******************


\titlerunning{Deep Coded Aperture Light Field Imaging}
\author{Mantang Guo\inst{1}\and
Junhui Hou\inst{1}\and
Jing Jin\inst{1}\and
Jie Chen\inst{2}\and
Lap-Pui Chau\inst{3}
}
\authorrunning{M. Guo et al.}
\institute{Department of Computer Science, City University of Hong Kong\and
Department of Computer Science, Hong Kong Baptist University\and
School of Electrical and Electronics Engineering, Nanyang Technological University\\
\email{mantanguo2-c@my.cityu.edu.hk, jh.hou@cityu.edu.hk (Corresponding author), jingjin25-c@my.cityu.edu.hk, chenjie@comp.hkbu.edu.hk, elpchau@ntu.edu.sg}}

%******************
\maketitle

\begin{abstract}
Coded aperture is a promising approach for capturing the 4-D light field (LF), in which the 4-D data are compressively modulated into 2-D coded measurements that are further decoded by reconstruction algorithms. The bottleneck lies in  the reconstruction algorithms, resulting in rather limited reconstruction quality. To tackle this challenge, we propose a novel learning-based framework for the reconstruction of high-quality LFs from acquisitions via learned coded apertures. The proposed method incorporates the measurement observation into the deep learning framework elegantly to avoid relying entirely on data-driven priors for LF reconstruction. Specifically, we first formulate the compressive LF reconstruction as an inverse problem with an implicit regularization term. Then, we construct the regularization term with an efficient deep spatial-angular convolutional sub-network to comprehensively explore the signal distribution free from the limited representation ability and inefficiency of deterministic mathematical modeling. Experimental results show that the reconstructed LFs not only achieve much higher PSNR/SSIM but also preserve the LF parallax structure better, compared with state-of-the-art methods on both real and synthetic LF benchmarks. In addition, experiments show that our method is efficient and robust to noise, which is an essential advantage for a real camera system. The code is publicly available at \url{https://github.com/angmt2008/LFCA}. 
\keywords{Light Field, Coded Aperture, Deep Learning, Regularization, Observation Model}
\end{abstract}

\section{Introduction}\label{sec:introduction}
Owing to multi-view and depth information embedded in 4-D light fields (LFs), a large variety of LF based applications have emerged, e.g., image post-refocusing\cite{ng2006digital}, 3-D reconstruction \cite{wang2016depth}, saliency detection\cite{li2014saliency}, view synthesis\cite{kalantari2016learning,wing2018fast,jin2020learning,guo2018dense,zhu2019revisiting}. Different from earlier LF capturing approaches, i.e., camera gantry\cite{levoy1996light} and camera array\cite{wilburn2005high}, portable micro-lens array-based LF cameras\cite{Lytro,RayTrix} are more convenient and cost-effective for capturing a dense LF. By using a micro-lens array placed between the main lens and image sensor, the micro-lens array-based camera records the spatial and angular information of light rays into a multiplexing sensor with only a single shot. However, due to the limited sensor resolution, the projection of the 4-D LF into a 2-D image leads to an inevitable trade-off between spatial and angular resolution. Besides, the large amount of data of captured LFs poses a great challenge to storage and transmission.

To preserve the LF resolution and simultaneously reduce the data size of captured LFs, based on a traditional camera, the coded aperture camera was designed, which modulates light rays through the main lens into one or multiple coded measurements with the same size as that of the image sensor. Then, by employing LF reconstruction algorithms, a full 4-D LF can be generated from coded measurements. Earlier LF reconstruction methods\cite{liang2008programmable,nagahara2010programmable,babacan2012compressive,marwah2013compressive,yagi2017pca,miandji2019unified}, which either require relatively many measurements or use dictionaries, are limited by the representation ability. Recent deep learning-based methods\cite{inagaki2018learning,gupta2017compressive,nabati2018fast} are able to reconstruct LFs from only a few measurements. However, these methods are purely data-driven without taking the special characteristics of LFs into account. That is, they employ networks for general purposes but not specifically designed for tackling the problem of LF reconstruction from coded measurements, e.g., the plain convolutional layers, the networks used in the fully convolutional network (FCN)\cite{long2015fully}, and the very deep convolutional network (VDSR)\cite{kim2016accurate}, and thus the reconstruction performance is still limited. 

 \begin{figure}[t]
 \centering
 \includegraphics[width=\linewidth]{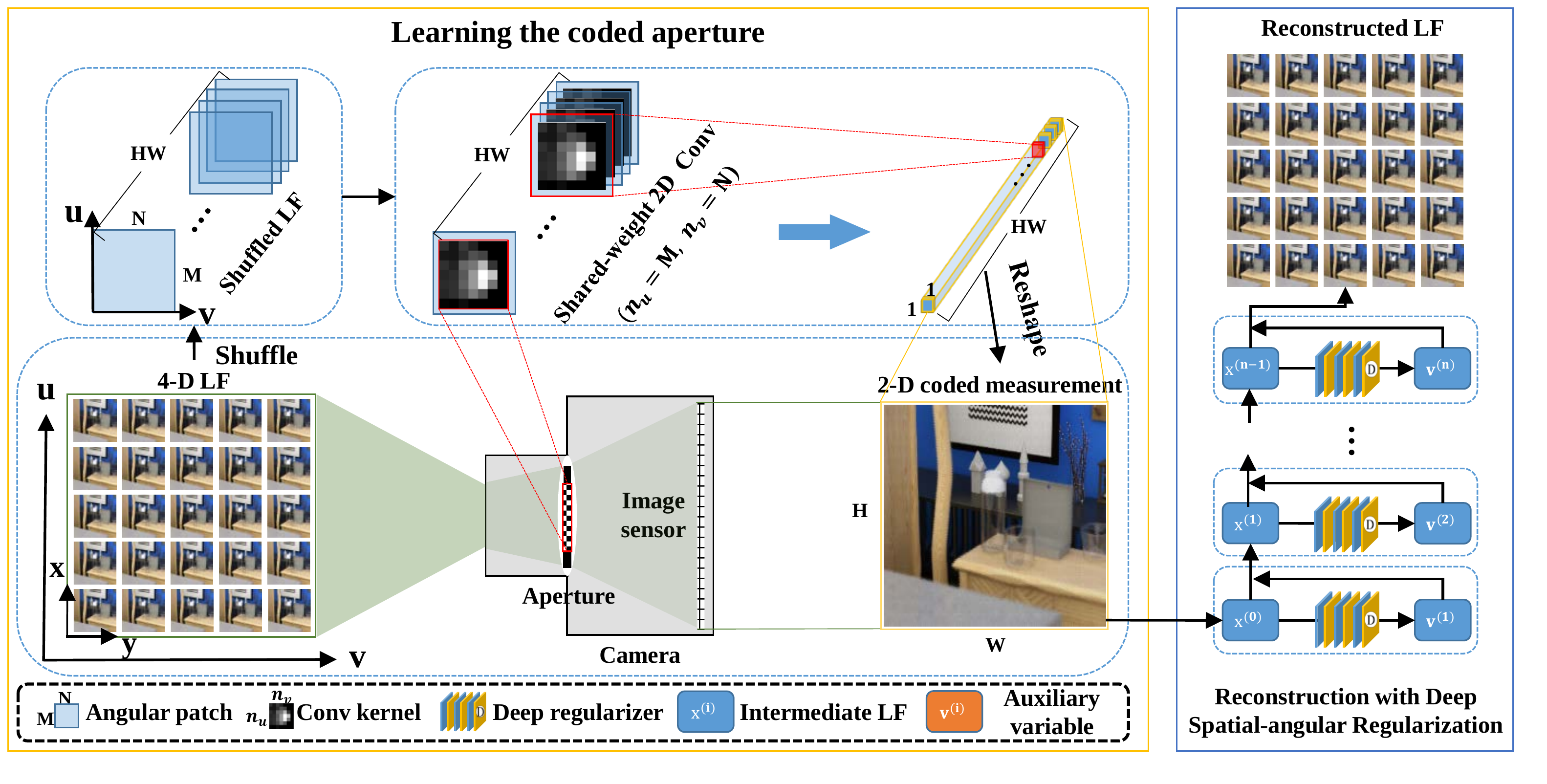}
 \caption{The pipeline of our deep learning-based compressive LF reconstruction over coded apertures. Our method elegantly incorporates the observation model of coded measurements into deep learning framework. The left side illustrates the acquisition of coded measurements by learning apertures, and the right side shows the reconstruction phase. More details of the reconstruction module are shown in Fig.~\ref{framework}}
 \label{pipeline}
 \end{figure}
 
In this paper, as shown in Fig.~\ref{pipeline}, we propose a novel deep learning-based framework to reconstruct high-quality LFs with measurements from adaptively learned coded apertures. First, the coded aperture is modeled and learned by a 2-D convolutional layer with specific configurations, denoted as acquisition layer in the method. Based on the observation model of the coded measurements, we formulate the LF reconstruction from measurements as an inverse problem with an implicit regularization term. Then, we construct the regularization term with a deep spatial-angular convolutional network instead of an deterministic mathematical modeling with a limited representation ability, such that the underlying complex structure can be comprehensively explored. Consequently, the LF reconstruction from coded measurements is solved by training the end-to-end network. Our LF capture and reconstruction method can breakthrough the limitations of conventional optimization method and simultaneously take full advantage of the strong representation ability of deep learning. Note that this paper is focused on developing a novel LF reconstruction method from coded measurements. Together with the experimentally verified robustness of our method against noise as well as the well constructed hardware platforms of coded aperture LF imaging in previous works\cite{liang2008programmable,nagahara2010programmable,marwah2013compressive,chen2015light,inagaki2018learning}, there is no technical barrier to implement the proposed method with a real camera system. 

\section{Related Work}\label{sec:related work}
Based on the inputs, we divide the existing LF reconstruction methods into two categories: sub-aperture image (SAI)-based reconstruction and coded measurement-based reconstruction, which will be reviewed as follows.
\subsection{LF Reconstruction from SAIs}
For SAI-based LF reconstruction, the input consists of a sparse set of SAIs belonging to a dense LF to be reconstructed. This kind of methods mainly investigate view synthesis to increase the angular resolution of a sparsely-sampled LF, which is always fixed into a regular pattern, e.g., four-corner SAIs\cite{kalantari2016learning}, borders-diagonal SAIs\cite{shi2014light}, SAI-pairs\cite{yoon2015learning}, or multiplane images\cite{zhou2018stereo,mildenhall2019local}. Specifically, Shi \textit{et al.}\cite{shi2014light} reconstructed a dense LF from fixed SAI-sampling patterns by exploiting the sparsity in Fourier domain. Yoon \textit{et al.}\cite{yoon2015learning} proposed a deep learning method to synthesize novel SAI between a pair or stack of SAIs. Besides, input SAIs obtained from different sampling patterns are processed separately by three sub-networks, which is inefficient for LF super-resolution. Kalantari \textit{et al.}\cite{kalantari2016learning} used four-corner SAIs as the input of their algorithm. The quality of reconstructed LFs is rather limited by using a series of hand-crafted features which are extracted from warped images. In addition, it fails to handle the input only with a single SAI as the hand-crafted features cannot be obtained. Such a weakness also exists in the method by Wu \textit{et al.}\cite{wu2017light}, which reconstructs a dense LF by carrying out super-resolution on angular dimension of each epipolar plane image (EPI) of the input LF. Yeung \textit{et al.}\cite{wing2018fast} proposed an end-to-end deep learning method to reconstruct a dense LF in a coarse-to-fine manner. However, the method\cite{wing2018fast} also cannot reconstruct a dense LF from few SAIs. Although the method proposed by Srinivasan \textit{et al.}\cite{srinivasan2017learning} can reconstruct a dense LF from a single SAI, it is only able to use information from the input SAI, and the coherence among SAIs of LF is lost. Moreover, the quality of reconstructed LF by the method relies heavily on the accuracy of depth estimation. 

\subsection{LF Reconstruction from Coded Measurements}
For coded measurement-based LF reconstruction, earlier methods\cite{liang2008programmable,nagahara2010programmable,babacan2012compressive,qu2017high} require relatively many exposures to reconstruct the entire LF. With the introduction of compressive sensing\cite{ashok2010compressive}, the dense LF can be reconstructed from coded measurements. Two challenges in LF compressive sensing are the design of sensing matrix for projecting LF into proper measurements and the algorithm for inversely reconstructing LF from measurements. Marwah \textit{et al.}\cite{marwah2013compressive} designed a sensing matrix by using conventional optimization method with an overcomplete dictionary. Then, the method in \cite{marwah2013compressive} formulates LF reconstruction from measurements as a basis pursuit denoise problem which is solved by conventional solvers. Chen \textit{et al.}\cite{chen2015light} constructed a dictionary based on perspective shifting of center view of an LF. Miandji \textit{et al.}\cite{miandji2019unified} proposed to aggregate multidimensional dictionary ensemble to encode and decode LFs efficiently in dictionary-based compressive sensing. However, the sensing matrix in these model-based methods has to be carefully designed to make the projection of LF as orthogonal as possible. Furthermore, a large scale of optimization techniques and training data are likely used to improve the representing ability of dictionaries. 

Recently, deep learning-based LF reconstruction methods were proposed, which design network architectures to infer LFs from coded measurements by training with a large amount of LF data. Based on the sensing matrix designed by \cite{marwah2013compressive}, Gupta \textit{et al.}\cite{gupta2017compressive} proposed a deep learning-based method for LF compressive sensing. Given coded measurements, the method in \cite{gupta2017compressive} employs two plain network branches to generate two coarse LFs and then fuses them together to generate the final LF. Nabati \textit{et al.}\cite{nabati2018fast} improved the sensing matrix which can modulate both color and angular information of an LF into the 2-D coded measurement. Then, the sensing matrix together with the coded measurements are fed into an FCN-based network to reconstruct a dense LF. However, since all these methods use a fixed sensing matrix to modulate LFs, the sensing process is not flexible enough to extract information from LFs. Furthermore, all these reconstruction networks are data-driven models without considering the signal reconstruction principle. Inagaki \textit{et al.}\cite{inagaki2018learning} adopted a $1\times1$ convolutional kernel to simulate the coded aperture process. Then, they employed two sequential sub-networks to reconstruct a dense LF from coded measurements. The first sub-network is constructed by a series of stacked convolutional layers while the second sub-network is a VDSR network\cite{kim2016accurate}. Although all angular information from different angular locations is selectively blended into coded measurements, the reconstruction quality is rather limited due to the plain network architecture.

\section{Proposed Method}\label{sec:proposed method}
As shown in Fig.~\ref{pipeline}, we model the compressive LF reconstruction over coded apertures within an end-to-end deep learning framework. Specifically, the measurements are obtained by modulating 4-D LFs through a learning-based coded aperture. For reconstruction, it is formulated as an inverse problem with a deep spatial-angular regularizer, which elegantly incorporates an LF degradation model into the deep learning framework. In what follows, we demonstrate each component in detail.
\subsection{Learning Coded Apertures}\label{sec:proposed method_1}
In the following, we just consider a single color channel of the LF image in RGB space for simplicity. The other two color channels will be processed in the same manner. The 4-D LF denoted as $\mathscr{L}(u,v,x,y)\in \mathbb{R}^{M\times N \times H\times W}$ can be represented with the two-parallel plane parameterization, where $\{(u,v)|u\in [1,M],v\in [1,N]\}$ and $\{(x,y)|x\in [1,H],y\in [1,W]\}$ are the angular and spatial coordinates, respectively. As shown in Fig.~\ref{pipeline}, incident light rays are modulated when passing through different aperture positions before converging at the image sensor. Then, the camera captures a 2-D coded measurement $\mathbf{L}_i(x,y)\in \mathbb{R}^{H\times W}$ of the LF. Specifically, the 2-D coded measurement can be formulated as
\begin{equation}
\label{e1}
\mathbf{L}_i(x,y)=\sum_{u=1}^{M}\sum_{v=1}^{N}a_i(u,v)\mathscr{L}(u,v,x,y),\\
\end{equation}
where $a_i(u,v)\in [0,1]$ is the transmittance at aperture position $(u,v)$ in the $i$-th capturing of LF. 

Based on the formulation, a coded measurement is the weighted summation of all SAIs. 
Our method simulates this process by a 2-D convolutional layer with specific configurations. The input of the layer is the entire 4-D LF while the output is 2-D coded measurements corresponding to the input LF. In our simulation, the convolutional operation is carried out on $u-v$ plane, i.e., the angular patch of the input LF, to fuse all $M\times N$ elements in the angular dimension into desired number of elements. We first shuffle the input LF to let the spatial dimension into one axis, i.e., the batch axis during training, to share a same kernel with all angular patches. By setting a proper kernel size, padding and strides, the kernel in the acquisition layer is able to fully cover or slide on the angular patch to obtain desired number of measurements. For example, to capture an LF with angular resolution $7\times 7$ into corresponding $1$, $2$, and $4$ measurements, we set the corresponding kernel size $n_{u}\times n_{v}$ to $7\times 7$, $7\times 6$ and $6\times 6$ with zero padding and one-pixel stride to produce the desired number of measurements. Besides, we limit the weights into $[0,1]$ and set bias to $0$ in the acquisition layer corresponding to the physical capturing process during training.

Note that the learned aperture can be realized by a typical programmable device, e.g., liquid crystal on silicon (LCoS) display, in a real camera implementation \cite{liang2008programmable,nagahara2010programmable,marwah2013compressive,chen2015light,inagaki2018learning}. Moreover, there is the color mask which provides an opportunity for modulating both color and angles of incident light rays\cite{nabati2018fast}. Our method can also simulate color LF coded aperture capturing by using three channels in the convolutional layer to modulate the R, G, B channels of the input LF simultaneously. The results are demonstrated in Section 4.

\subsection{Reconstruction with Deep Spatial-angular Regularization}
\subsubsection{The observation model.} Based on the sensing mechanism in Sec.~\ref{sec:proposed method_1}, the observation model of coded aperture measurements can be written as  
\begin{equation}
\label{e2}
\mathbf{l}=\mathbf{Ax}+\bm{\epsilon},
\end{equation}
where $\mathbf{l}=[\mathbf{l}_1; \mathbf{l}_2; ...; \mathbf{l}_k]\in\mathbb{R}^{kHW}$ is the set of $k$ measurements with $\mathbf{l}_i\in \mathbb{R}^{HW}$ ($i=1, 2, ..., k$) being vectorial representation of the $i$-th measurement $\mathbf{L}_i$, $\mathbf{x}\in \mathbb{R}^{HWMN}$ is the vectorial representation of the original LF to be reconstructed, 
$\mathbf{A}\in \mathbb{R}^{kHW\times HWMN}$ denotes the linear degradation/sensing matrix, and $\bm{\epsilon}\in \mathbb{R}^{kHW}$ denotes the additive noise.

\subsubsection{Prior-driven solution.}
Since recovering $\mathbf{x}$ from $\mathbf{l}$ in Eq.~(\ref{e2}) is an ill-posed inverse problem, regularization has to be introduced to constrain the solution space of $\mathbf{x}$. Thus, the problem of LF reconstruction from measurements can be generally cast as 
\begin{equation}
\label{e3}
\mathop{\min}\limits_{\mathbf{x}}\frac{1}{2}\left \| \mathbf{l}-\mathbf{Ax} \right \|_{2}^{2} +\lambda \mathcal{J}(\mathbf{x}),
\end{equation}
where $\mathcal{J}(\cdot)$ is the regularization term, and $\lambda$ is a positive penalty parameter to balance the two terms.
 
By introducing an auxiliary variable $\mathbf{v}\in\mathbb{R}^{HWMN}$, the optimization problem in Eq.~(\ref{e3}) can be decoupled into two sub-problems correspondingly for the data likelihood term and the regularization term \cite{venkatakrishnan2013plug,romano2017little}:
\begin{equation}
\label{e4}
\mathop{\min}\limits_{\mathbf{x},\mathbf{v}}\frac{1}{2}\left \| \mathbf{l}-\mathbf{Ax} \right \|_{2}^{2} +\lambda \mathcal{J}(\mathbf{v}), \quad s.t. \quad \mathbf{x}=\mathbf{v}.
\end{equation}
We further convert Eq.~(\ref{e4}) into an unconstrained problem by moving the equality constraint into the objective function as a penalty term, i.e., 
\begin{equation}
\label{e5}
\mathop{\min}\limits_{\mathbf{x},\mathbf{v}}\frac{1}{2}\left \| \mathbf{l}-\mathbf{Ax} \right \|_{2}^{2} +\eta\left \|\mathbf{x}-\mathbf{v}\right \|_{2}^{2}+\lambda \mathcal{J}(\mathbf{v}),
\end{equation}
where $\eta >0$ is a penalty parameter. Based on the half quadratic splitting method\cite{zhang2017learning,dong2018denoising}, the optimization problem in Eq.~(\ref{e5}) can be solved by alternatively solving the following two sub-problems until convergence:
\begin{equation}
\label{e6}
\left \{
\begin{aligned}
&\mathbf{x}^{(t+1)}=\mathop{\argmin}\limits_{\mathbf{x}}\frac{1}{2}\left \| \mathbf{l}-\mathbf{Ax} \right \|_{2}^{2} +\eta\left \|\mathbf{x}-\mathbf{v}^{(t)}\right \|_{2}^{2},\\
&\mathbf{v}^{(t+1)}=\mathop{\argmin}\limits_{\mathbf{v}}\eta \left \| \mathbf{x}^{(t+1)}-\mathbf{v} \right \|_{2}^{2} +\lambda \mathcal{J}(\mathbf{x}^{(t+1)}),
\end{aligned}
\right.
\end{equation}
where $t$ is the iteration index. For the $\mathbf{x}$-subproblem in Eq.~(\ref{e6}), it can be computed with a single step of gradient descent for an inexact solution:
\begin{equation}
\label{e7}
	\mathbf{x}^{(t+1)}=\mathbf{x}^{(t)}-\delta [\mathbf{A}^\textsf{T}(\mathbf{A}\mathbf{x}^{(t)}-\mathbf{l})+\eta(\mathbf{x}^{(t)}-\mathbf{v}^{(t)})],\\
\end{equation}
where $\delta>0$ is the parameter controlling the step size. With regard to the $\mathbf{v}$-subproblem, the solution is the proximity term of $\mathcal{J}(\mathbf{v})$ at the point, i.e., 
\begin{equation}
\label{e8}
\mathbf{v}^{(t+1)}=\mathcal{D}(\mathbf{x}^{(t+1)}),
\end{equation}
where $\mathcal{D}(\cdot)$ denotes the proximal operator with respect to a typical regularization $\mathcal{J}(\cdot)$.

 \begin{figure}[t]
 \centering
 \includegraphics[width=\textwidth]{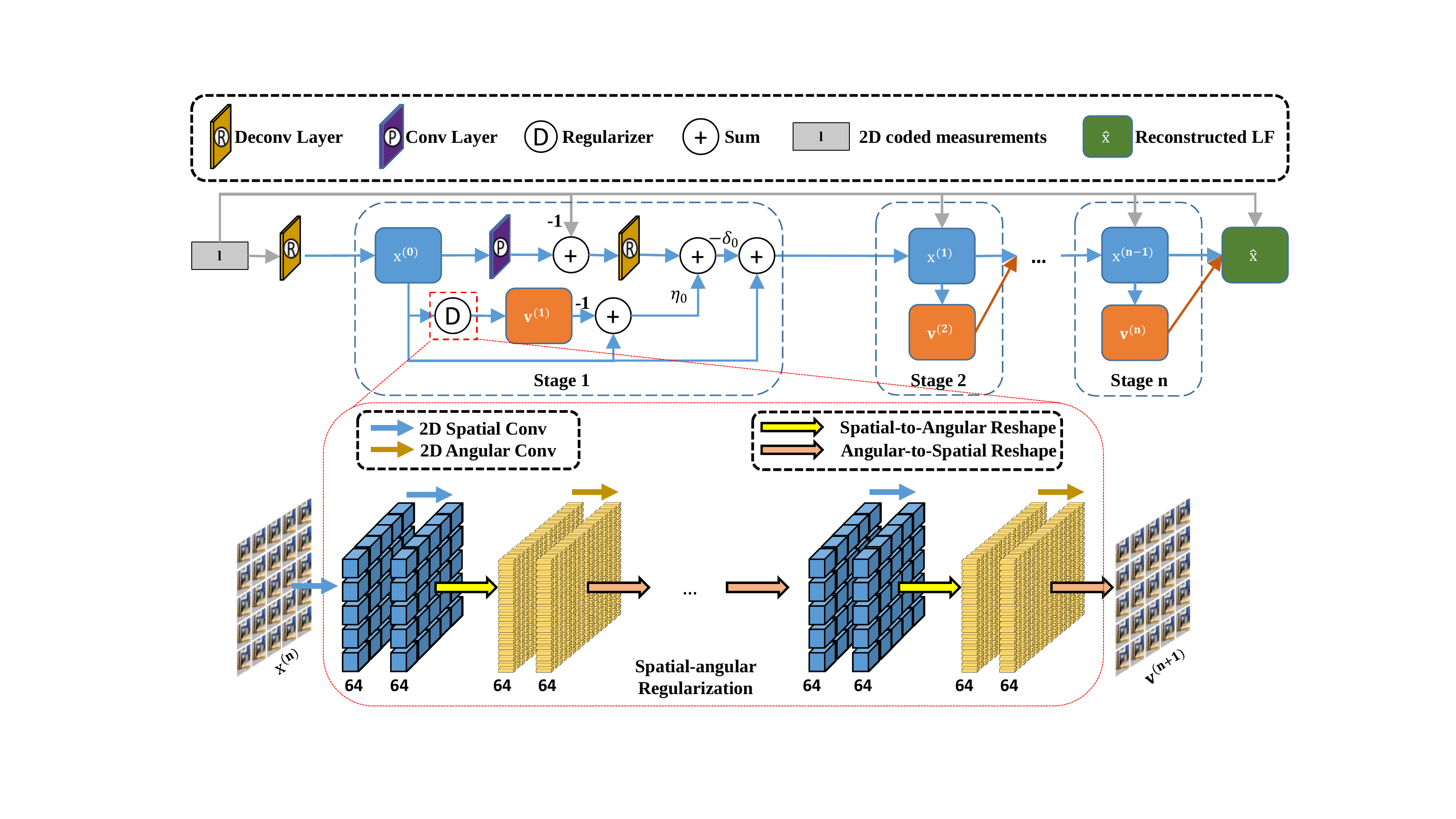}
 \caption{The architectures of the proposed iterative framework and the deep spatial-angular regularization sub-network}
 \label{framework}
 \end{figure}
 
\subsubsection{Deep spatial-angular regularization.}
Due to the high-dimensional property and complex geometry structure in LFs, it is difficult to use an explicit regularization term in Eq.~(\ref{e8}), which commonly has a limited representation ability, for comprehensively exploring the underlying distribution. To this end, as shown in Fig.~\ref{framework}, we adopt a deep implicit regularization which is constructed by computationally-efficient spatial-angular separable (SAS) convolutional layers\cite{yeung2018light,wing2018fast}. The SAS convolution is able to thoroughly detect the dimensional correlations of a 4-D pixel in the LF by alternatively conducting 2-D convolutional operations on spatial and angular planes. Besides, SAS convolution does not significantly increase the number of parameters compared against the 4-D convolution. Furthermore, linear transformations which are implemented by matrices in conventional optimization algorithms, i.e., Eqs.~(\ref{e7}) and (\ref{e8}), can be replaced by convolutional layers or networks without impairing the convergence property\cite{zhang2018ista,sun2016deep,dong2018denoising}.
Since the degradation matrix $\mathbf{A}$ and its transpose $\mathbf{A}^{\textsf{T}}$ in Eq.~(\ref{e7}) are linear projections, we can correspondingly replace them with a convolutional layer as mentioned in Sec.~\ref{sec:proposed method_1} and a corresponding deconvolutional layer for inverse projection. Furthermore, in order to preserve the linear property of the transformations, all these layers are not followed by activation functions or bias units. The projection $\mathcal{P}(\cdot)$ conducted by the the convolutional layer can be regarded as a linear mapping function which projects $\mathbf{x}$ to $\mathbf{l}$. On the contrary, the inverse projection $\mathcal{R}(\cdot)$ conducted by a deconvolutional layer is regarded as a linear function which projects $\mathbf{l}$ to $\mathbf{x}$, i.e., 
\begin{equation}
\label{e9}
\mathbf{l}=\mathcal{P}(\mathbf{x}), \mathbf{x}=\mathcal{R}(\mathbf{l}).
\end{equation}
Thus, Eq.~(\ref{e8}) and Eq.~(\ref{e7}) can be respectively rewritten as
\begin{equation}
\label{e10}
\left \{
\begin{aligned}
&\mathbf{v}^{(t+1)}=\mathcal{D}(\mathbf{x}^{(t)},\theta_{d}^{t}),\\
&\mathbf{x}^{(t+1)}=\mathbf{x}^{(t)}-\delta_{t} [\mathcal{R}( \mathcal{P}(\mathbf{x}^{(t)},\theta_{p}^{t})-\mathbf{l},\theta_{r}^{t})+\eta_{t}(\mathbf{x}^{(t)}-\mathbf{v}^{(t+1)})],
\end{aligned}
\right .
\end{equation}
where $\theta_{r}, \theta_{p}$, and $\theta_{d}$ are the network parameters which will be learned by the backpropagation algorithm during the training process. In order to enhance the representing ability of the network, parameters in each iterative stage are independently learned without being inherited. 

Our iterative framework for reconstructing a 4-D LF from 2-D coded measurements is demonstrated in Fig.~\ref{framework}. Given 2-D coded measurements $\mathbf{l}$, we first use an inverse projection $\mathcal{R}(\mathbf{l},\theta_{r}^{0})$ to produce an initialization $\mathbf{x}^{(0)}$ for the reconstructed LF. After being processed by the deep regularization $\mathcal{D}(\mathbf{x}^{(0)},\theta_{d}^{0})$, an optimized LF $\mathbf{x}^{(1)}$ which is better than that from last iterative stage is estimated. Such an iterative stage repeats $n$ times to gradually generate a final reconstructed LF.

\subsection{Training and Implementation Details}
\subsubsection{Training strategy and parameter setting.} 
In our method, the acquisition layer is a 2-D convolutional layer without activation function or bias units for the simulation of coded aperture modulation. The parameters, i.e., the kernel size, padding and stride, in the layer can be flexibly set before training for different simulation tasks. Correspondingly, the parameters in the deconvolutional layer are the same as those in the convolutional layer. It is to ensure that the output and input size of the deconvolutional layer are respectively the same as the input and output size of the convolutional layer. The loss function is the $\ell_1$ loss between reconstructed LF and the ground-truth LF. In the regularization sub-network, the kernel size in both spatial and angular convolutional layer is $3\times 3$. The number of feature maps in each layer is $64$. The output of each convolutional layer is mapped by a ReLU activation function. Besides, the number of SAS convolutional layers is set to $9$ according to our ablation studies in Sec.~\ref{sec:experiments_2}. At the training stage, patches of spatial size $32\times 32$ were randomly cropped from LFs contained in the training set. The batch size was set to $5$. In order to increase the number of training samples, we randomly cropped 4-D LF patch with size of $M\times N\times 32\times 32$ from each LF image in the training datasets. The learning rate was initially set to  $1e-4$  and reduced to $1e-5$ when the loss stopped decreasing. We chose Adam\cite{kingma2014adam} as the optimizer with $\beta_{1}=0.9$ and  $\beta_{2}=0.999$. Our framework was implemented with PyTorch.

\subsubsection{Datasets.} The training dataset contains both synthetic and real-world LF images. Specifically, there are $100$ real-world LF images of size $7\times 7\times 376\times 541$ from Kalantari Lytro\cite{kalantari2016learning}, $22$ synthetic LF images of size $5\times 5\times 512\times 512$ from  HCI\cite{honauer2016benchmark}, and $33$ synthetic LF images of size $5\times 5\times 512\times 512$ from Inria\cite{shi2019framework}. The test set contains $30$ LF images from Kalantari Lytro\cite{kalantari2016learning}, $2$ LF images from HCI\cite{honauer2016benchmark} and $4$ LF images from Inria\cite{shi2019framework}. Please refer to the supplementary material for more details of the employed training and testing data.

\section{Experiments}\label{sec:experiments}
In this section, we evaluated the proposed LF reconstruction method by comparing our method with three state-of-the-art methods, followed by a series of comprehensive ablation studies.  
\subsection{Comparison with State-of-the-art Methods}
We compared with one state-of-the-art deep learning-based LF reconstruction method from coded aperture measurements, i.e., Inagaki \textit{et al.}\cite{inagaki2018learning}, and two deep learning-based methods from sparsely sampled SAIs, i.e., Kalantari \textit{et al.}\cite{kalantari2016learning} and Yeung \textit{et al.}\cite{wing2018fast}. According to Inagaki \textit{et al.}\cite{inagaki2018learning}, the performance of traditional compressive sensing-based methods is far below that of deep learning-based ones. Here we omitted the comparison with those methods. Specifically, the detailed experimental settings are listed as follows for fair comparisons: 

\begin{itemize}[noitemsep, topsep=0pt]
\item [$\bullet$] 
all the networks under comparison were re-trained with the same datasets using their source codes with suggested parameters;
\item[$\bullet$]
we conducted three tasks, i.e., $1\rightarrow49$, $2\rightarrow49$ and $4\rightarrow49$ ($i\rightarrow j$ denotes using $i$  measurements/SAIs to reconstruct an LF with angular resolution $j$) on real-world dataset, while one task $2\rightarrow25$ on synthetic dataset;
\item [$\bullet$]
we used a same single-channel kernel in our acquisition layer to modulate three color channels of LF, denoted as Ours (Single), for fairly comparing with  
Inagaki \textit{et al.}\cite{inagaki2018learning}. According to the analysis in Sec.~\ref{sec:proposed method_1}, a three-channel kernel was also trained in another model, denoted as Ours (Multiple);
\item [$\bullet$]
Kalantari \textit{et al.}\cite{kalantari2016learning} cannot handle the task $1\rightarrow49$ or $1\rightarrow25$ since the hand-craft features cannot be calculated in their algorithm. Besides, Kalantari \textit{et al.}\cite{kalantari2016learning} can accept flexible and irregular input patterns. As suggested in \cite{jin2019deep}, we re-trained it with choosing the optimal input patterns for comparison, i.e., the angular coordinates are correspondingly $(4,2)$ and $(4,6)$ for task $2\rightarrow 49$, while $(2,5)$, $(3,2)$, $(5,6)$ and $(6,3)$ for task $4\rightarrow 49$;
\item [$\bullet$]
and Yeung \textit{et al.}\cite{wing2018fast} requires the input SAIs with a regular pattern, and it cannot reconstruct an LF from only 1 or 2 SAIs. We thus re-trained it only on the task $4\rightarrow 49$ with the four-corner SAIs as inputs.
\end{itemize}

 \begin{figure*}[t]
 \centering
 \captionsetup{skip=0pt}
  \subfigure{\includegraphics[width=0.35\textwidth]{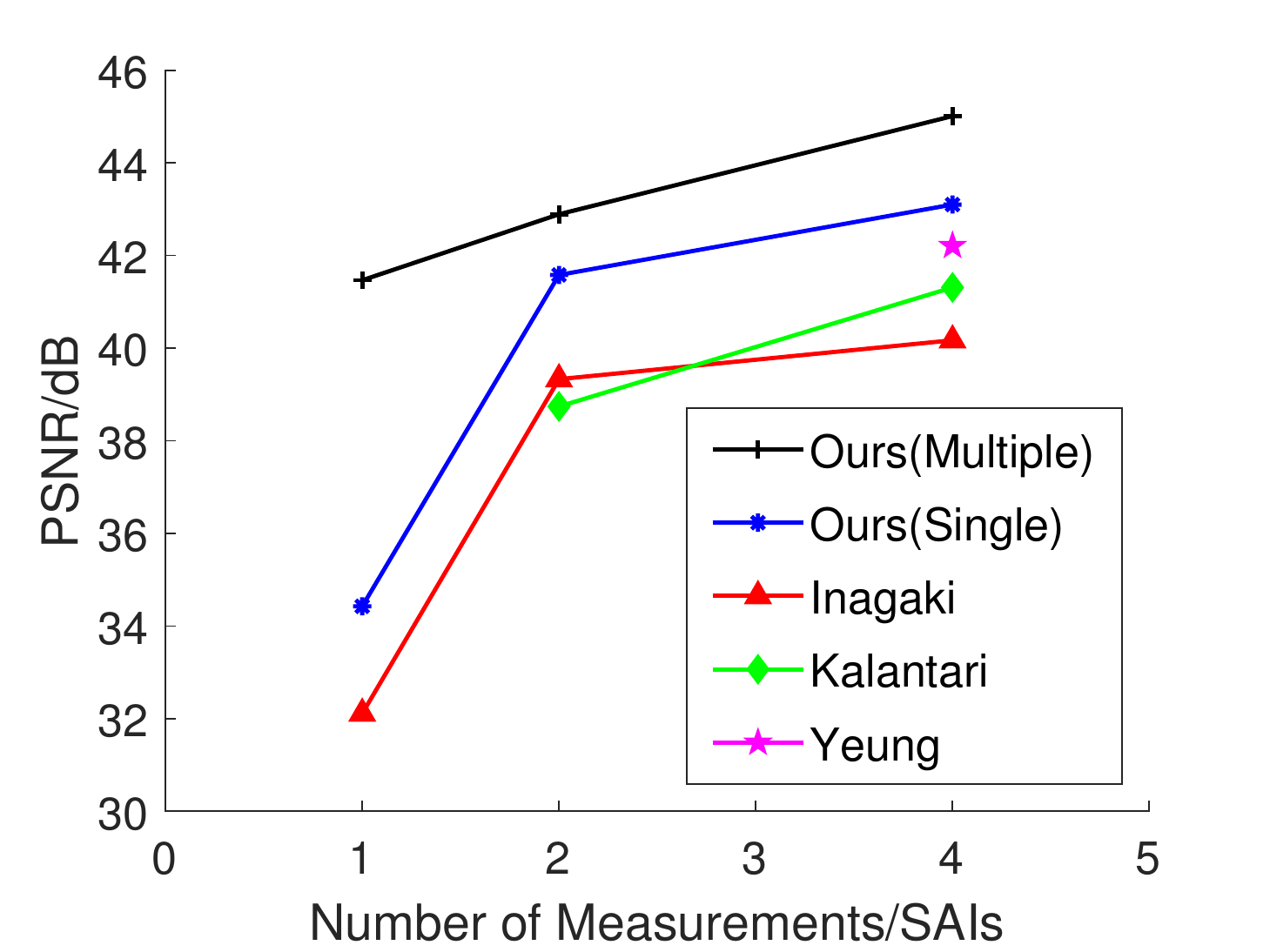}} 
  \subfigure{\includegraphics[width=0.35\textwidth]{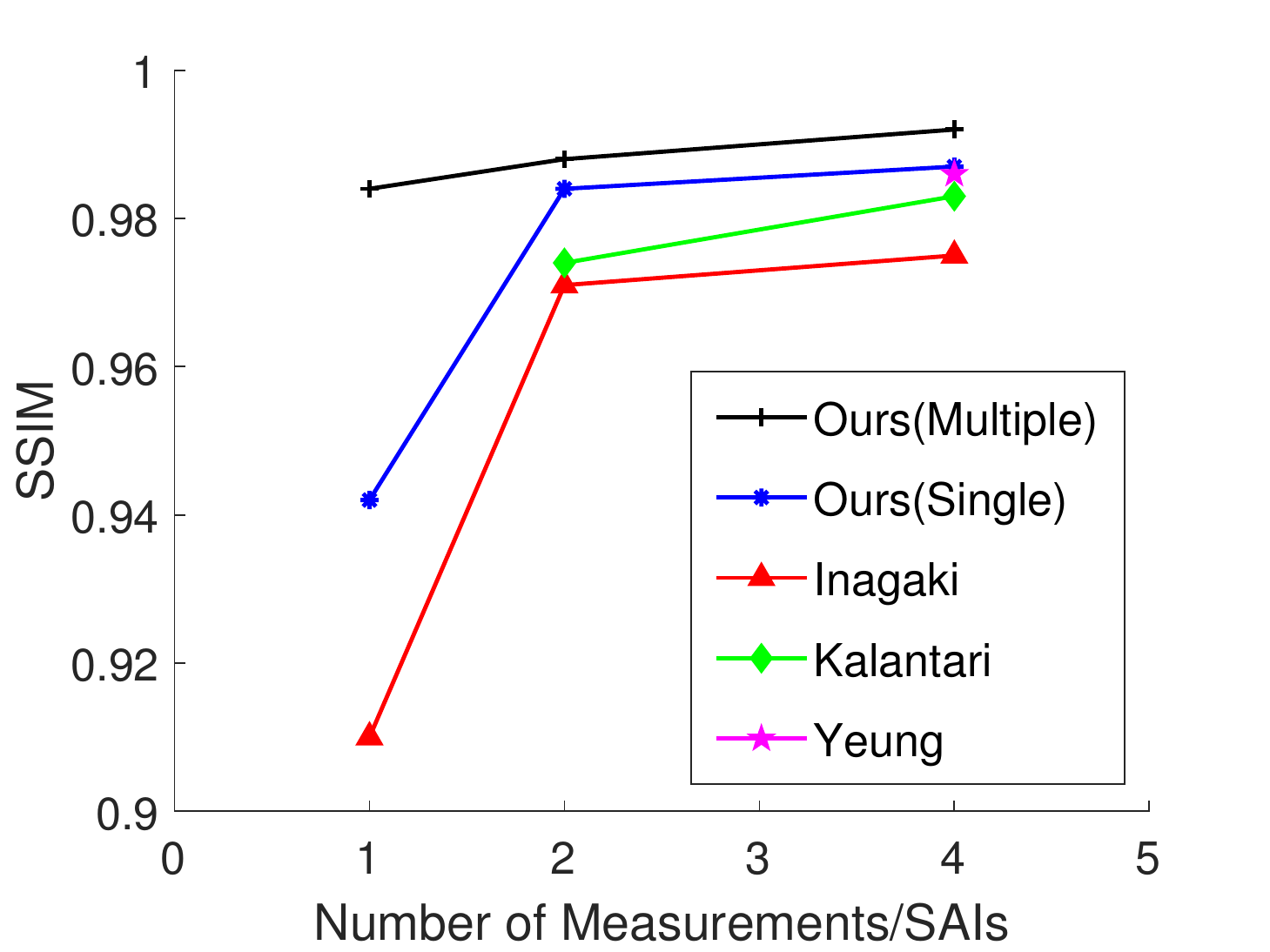}} \\
 \caption{The quantitative comparisons of all methods on various reconstruction tasks: $1\rightarrow 49$, $2\rightarrow 49$, and $4\rightarrow 49$. Here the PSNR and SSIM values refer to the average of all $30$ LFs contained in the test set from Kalantari Lytro\cite{kalantari2016learning}. See the supplementary material for the PSNR/SSIM of each LF image}
 \label{vwise_ps}
 \end{figure*}

\begin{figure*}[t]
\captionsetup{skip=0pt}
\centering
\subfigure[]{\includegraphics[width=0.16\textwidth]{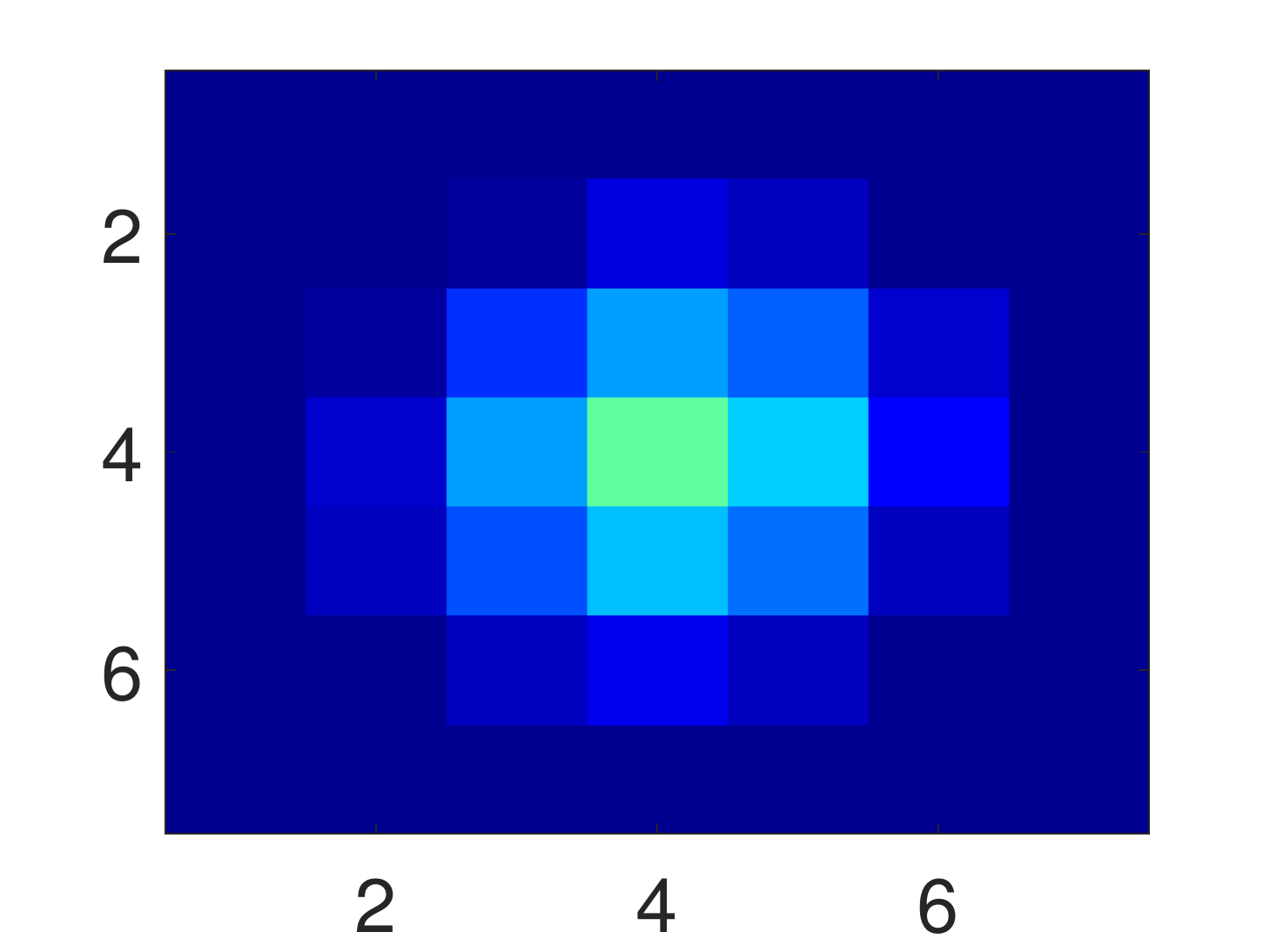}}
\subfigure[]{\includegraphics[width=0.16\textwidth]{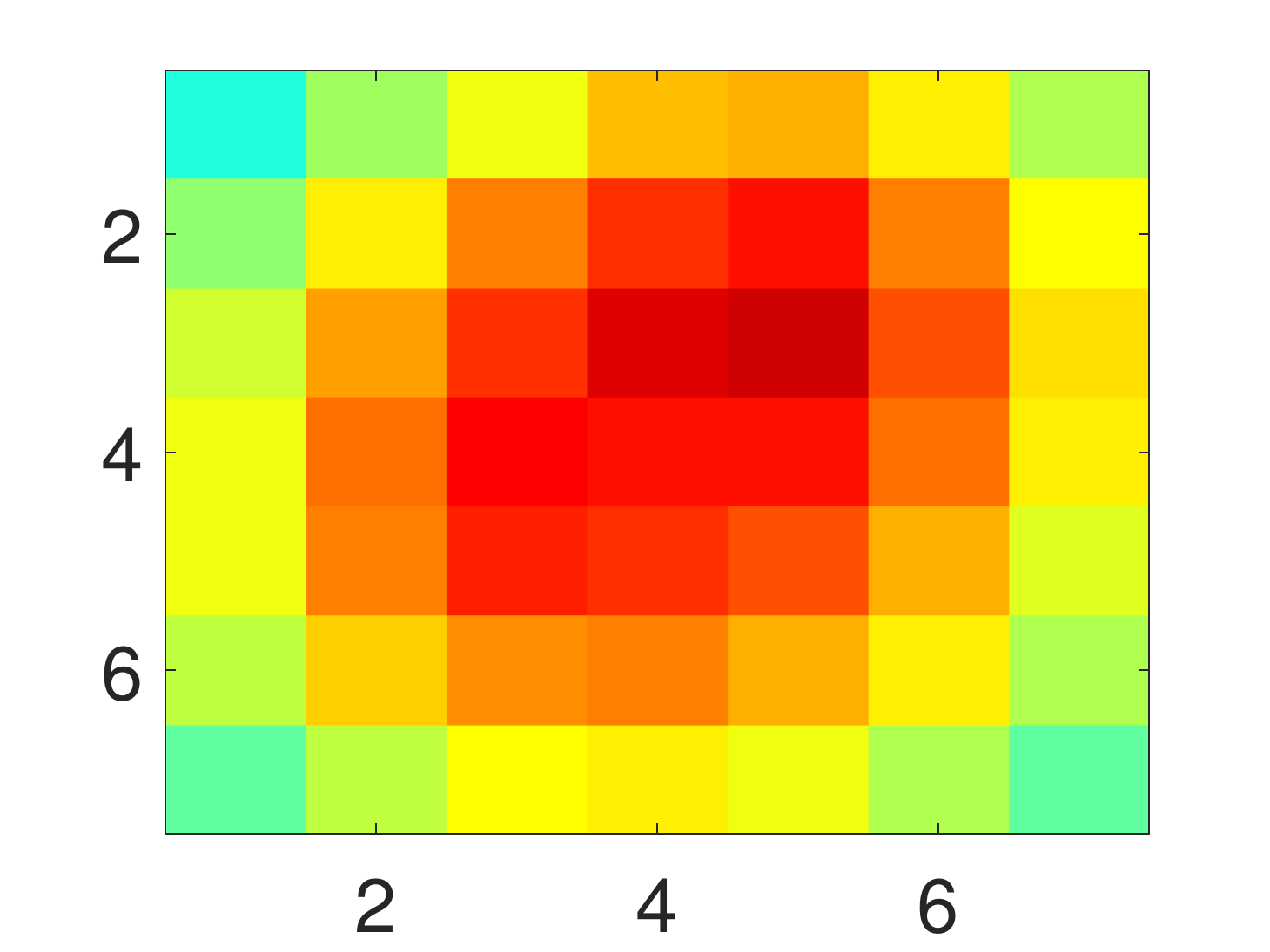}} 
\subfigure[]{\includegraphics[width=0.16\textwidth]{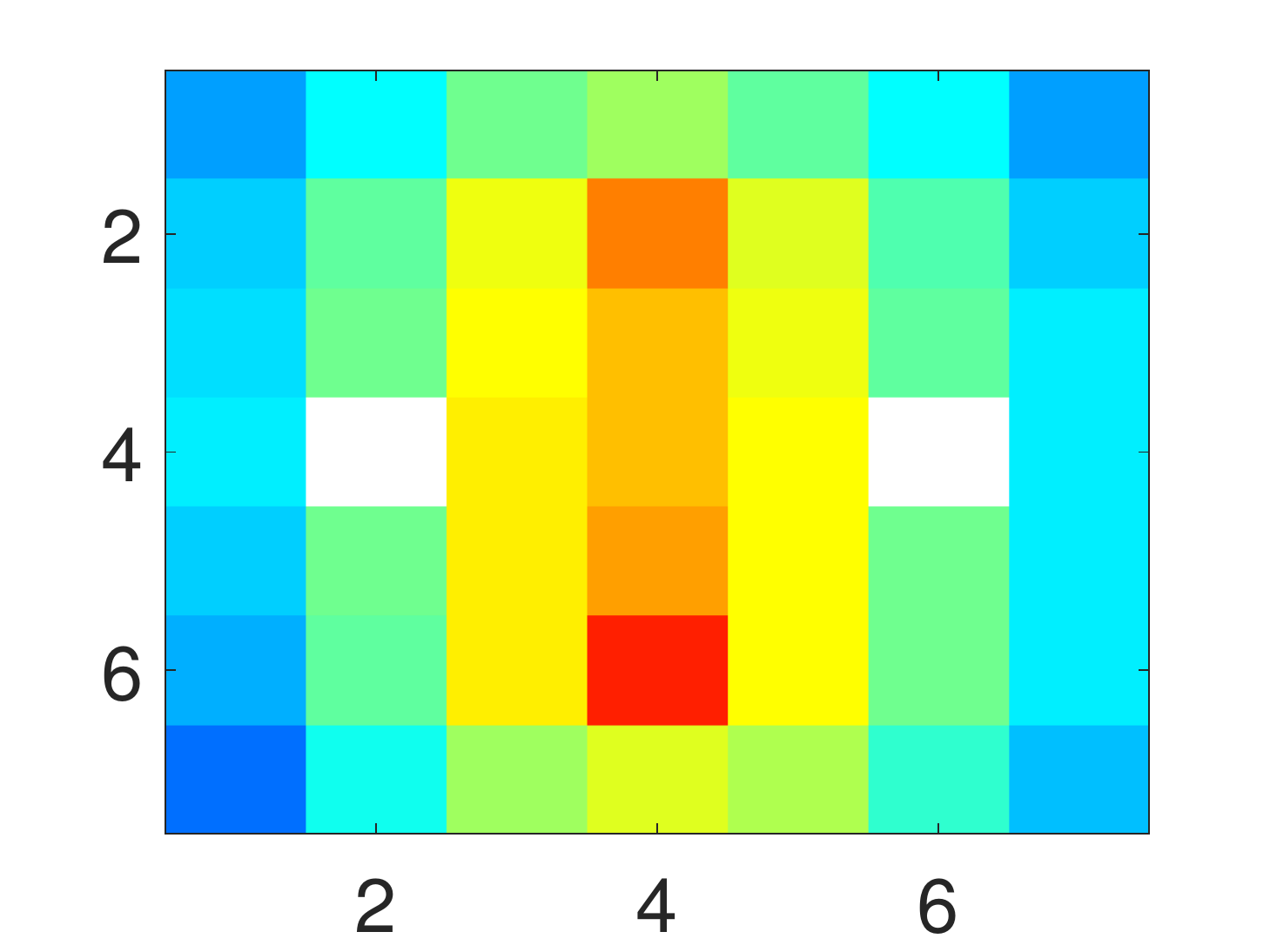}}
\subfigure[]{\includegraphics[width=0.16\textwidth]{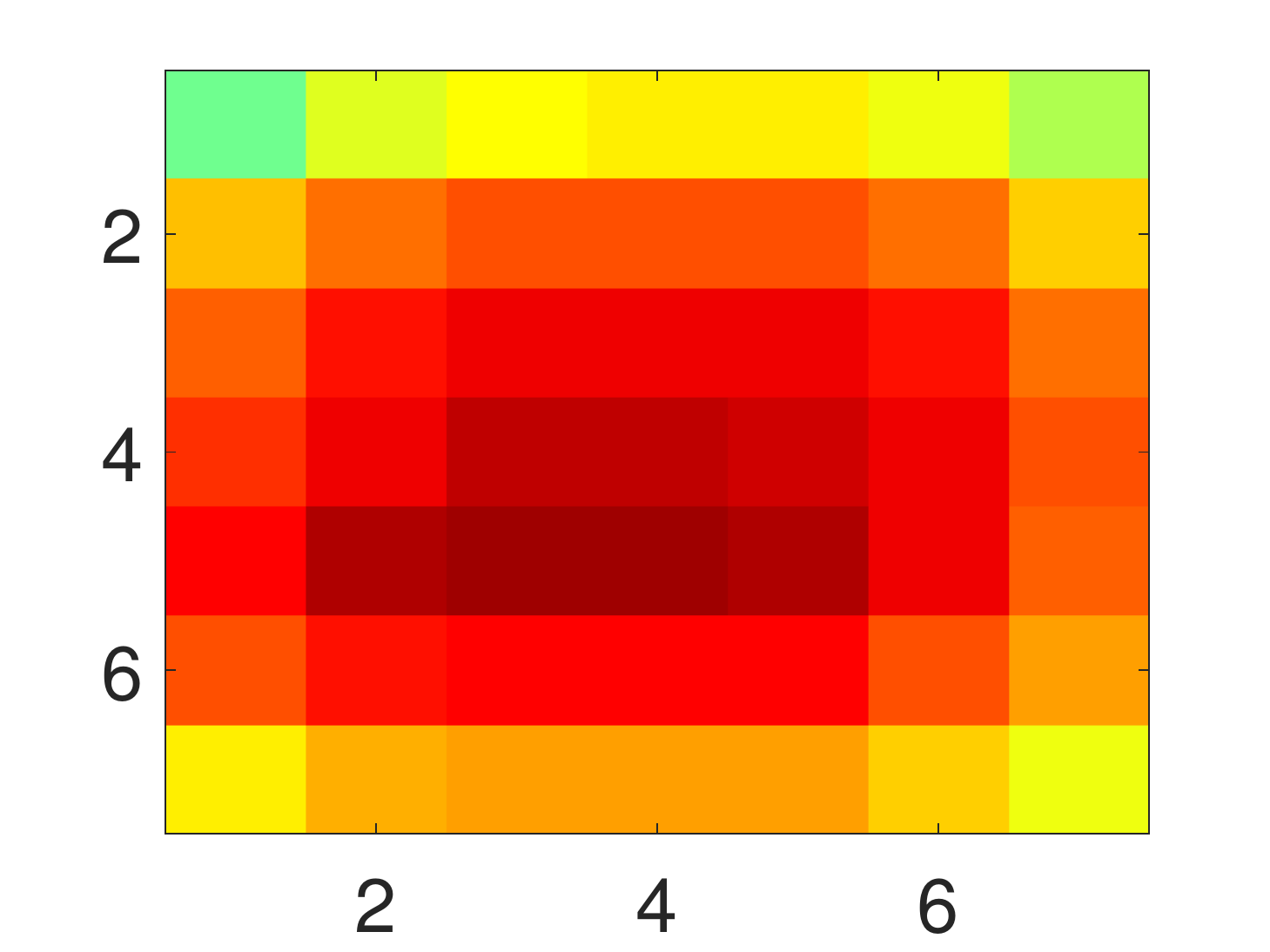}}
\subfigure[]{\includegraphics[width=0.16\textwidth]{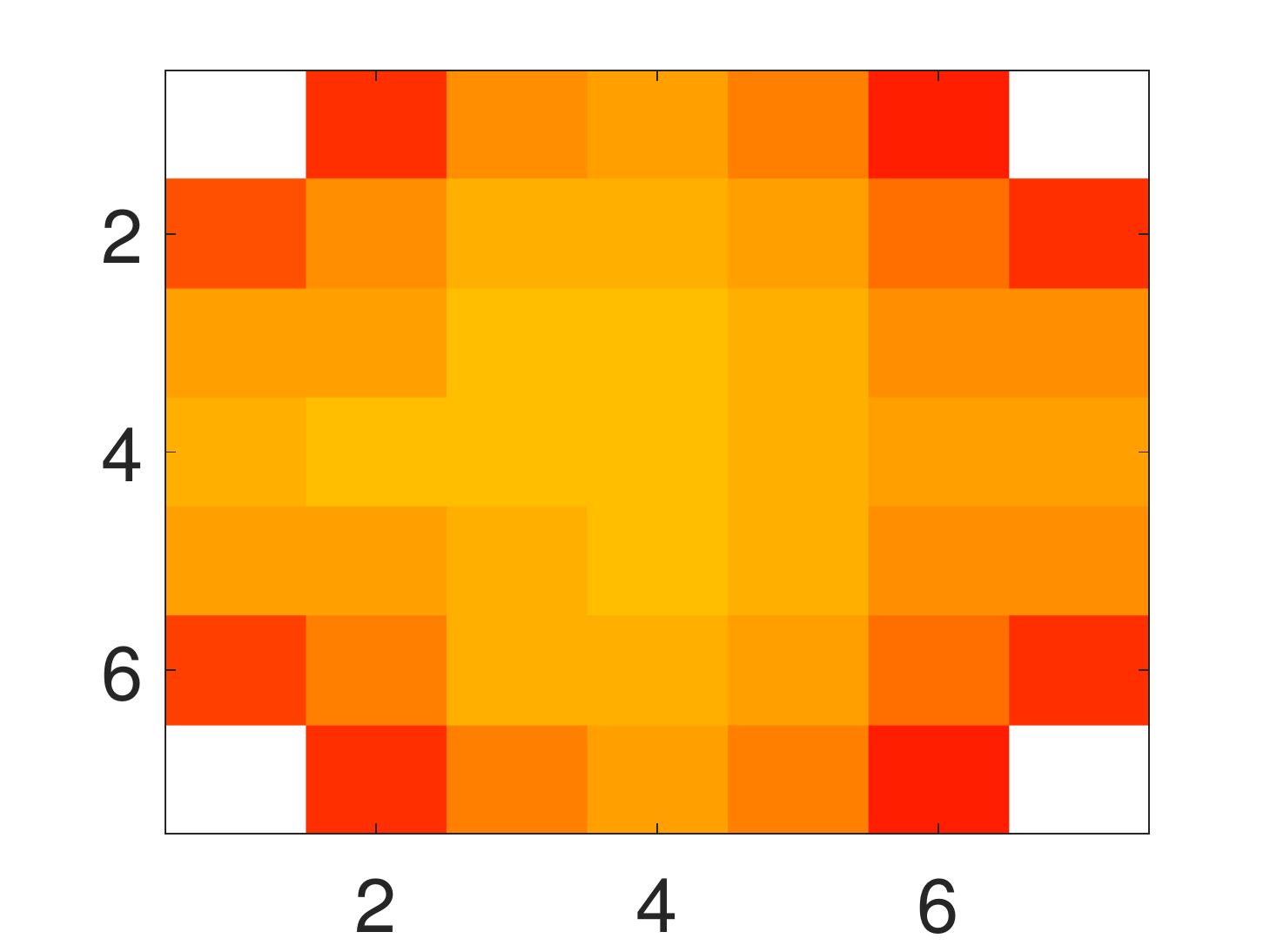}}   
\subfigure[]{\includegraphics[width=0.16\textwidth]{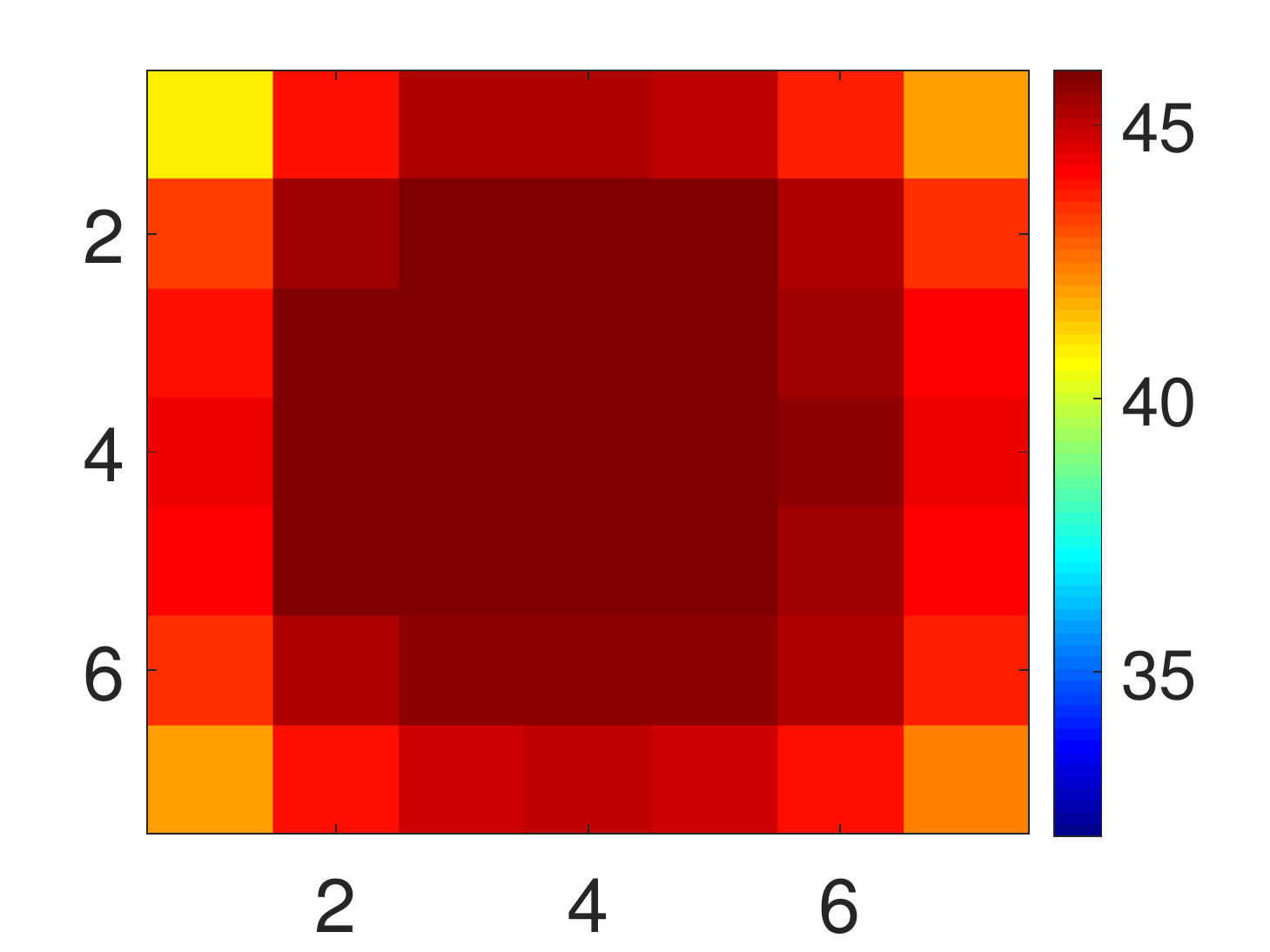}} \\
\caption{The average PSNR at each angular position of reconstructed LFs from different methods. The white blocks denote the input SAI positions. From left to right: (a) Inagaki $(1)$\cite{inagaki2018learning}, (b) Ours (Single) $(1)$, (c) Kalantari $(2)$\cite{kalantari2016learning}, (d) Ours (Single) $(2)$, (e) Yeung $(4)$\cite{wing2018fast}, (f) Ours (Single) $(4)$. The digits in brackets are the number of input measurements/SAIs for each method}
\label{viewpsnr}
\end{figure*}

\subsubsection{Quantitative comparison.}
We calculated the average PSNR and SSIM between the reconstructed LFs and ground-truth ones in RGB space to conduct quantitative comparisons of different methods, as shown in Fig.~\ref{vwise_ps}. Besides, the average PSNR at each SAI position of the reconstructed LFs from these methods are shown in Fig.~\ref{viewpsnr}. From Figs.~\ref{vwise_ps} and  \ref{viewpsnr}, we can draw the following conclusions:

\begin{itemize}[noitemsep, topsep=0pt]
\item [$\bullet$]
Ours (Multiple) has better performance than Ours (Single) under all tasks. The possible reason is that the three color channels may have different distributions, and the color channel-tailored aperture can adapt to each channel better than that a common one for three channels;
\item [$\bullet$]
both Ours (Single) and Ours (Multiple) consistently outperform the coded aperture method Inagaki \textit{et al.}\cite{inagaki2018learning} under all tasks. The reason is that Inagaki \textit{et al.}\cite{inagaki2018learning} employs network architecture which is not specifically designed for LF reconstruction, while our algorithm incorporates the observation model for measurements into the deep learning framework elegantly to avoid relying entirely on data-driven priors;
\item [$\bullet$]
and our method can preserve a high-quality reconstruction at most aperture positions. Conversely, the quality of SAI in Kalantari \textit{et al.}\cite{kalantari2016learning} and Yeung \textit{et al.}\cite{wing2018fast} declines along with the increase of the distance from input SAIs. The possible reason is that they can only extract features from input SAIs, while our method can adaptively condense and utilize information from all aperture positions through measurements.
\end{itemize}

\subsubsection{Visual comparison of reconstructed LFs.}
\begin{figure}[!t]
\centering
\includegraphics[width=0.9\textwidth]{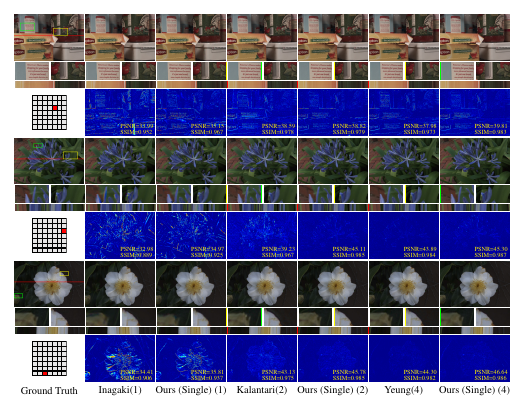}
\caption{Visual comparisons of all methods over real LF images under various reconstruction tasks: $1\rightarrow 49$, $2\rightarrow 49$ and $4\rightarrow 49$. The error maps are calculated in gray-scale space. The digits in the brackets are the number of input measurements/SAIs for the methods}
\label{vwise}
\end{figure}

\begin{figure}[t]
\centering
\includegraphics[width=0.7\textwidth]{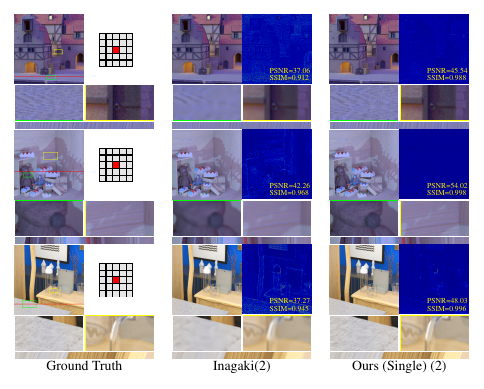}
\caption{Visual comparisons of our method against Inagaki\cite{inagaki2018learning} over synthetic LF data under the task $2\rightarrow 25$}
\label{syn}
\end{figure}

The visual comparisons of reconstructed LFs from all methods are shown in Figs.~\ref{vwise} and \ref{syn}, where it can be observed that our method can produce better details than other methods\cite{inagaki2018learning,kalantari2016learning,wing2018fast} under all tasks. For Kalantari \textit{et al.}\cite{kalantari2016learning} and Yeung \textit{et al.}\cite{wing2018fast}, the blurred artifacts and ghost effects exist at occlusion boundaries and high-frequency regions. Additionally, our method can produce better details than Inagaki \textit{et al.}\cite{inagaki2018learning} on synthetic LFs with large disparities. The reason is that the deep spatial-angular regularization in our method can fully exploit dimensional correlations in LFs to achieve a better reconstruction on datasets with large disparities.

\subsubsection{Comparison of the LF parallax structure.} 
Since the parallax structure among SAIs of an LF is the most valuable information, we conducted experiments to evaluate such a structure embedded in the reconstructed LFs from different methods. First, we compared the EPIs at the bottom of each subfigure of Figs. ~\ref{vwise} and \ref{syn}, where it can be seen that the EPIs
from our method show clearer and more consistent straight lines compared with those from other methods. Moreover, we conducted more investigations on this important issue. However, due to the lack of standard evaluation metrics, we chose the following two metrics to achieve the target:

\begin{figure}[!t]
\centering
\includegraphics[width=0.83\textwidth]{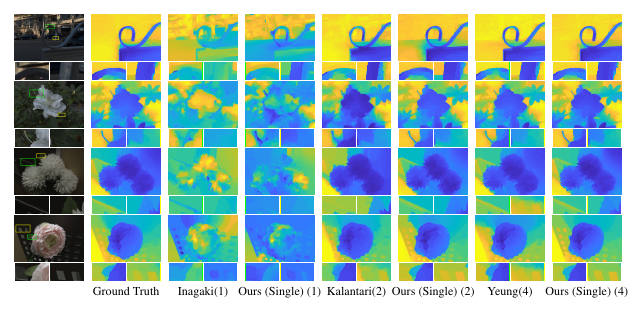}
\caption{Comparisons of depth maps estimated with reconstructed LFs by different methods under tasks $1\rightarrow 49$, $2\rightarrow 49$ and $4\rightarrow 49$. The digits in the brackets are the number of input measurements/SAIs for the methods}
\label{depth}
\end{figure}

(1). It is expected that with a typical LF depth estimation method, the estimated depth maps from LFs with better parallax structure will be more accurate and closer to those from ground-truth LFs. Thus, We applied the same depth estimation algorithm\cite{chen2018accurate} on the reconstructed LFs by different methods and the ground-truth ones. The visualized depth maps illustrated in Fig.~\ref{depth} show that  our method can preserve sharper edges at the occlusion boundaries, which are most similar to those of ground truth, which demonstrates the advantage of our method on preserving the LF parallax structure indirectly.

(2). Considering that the line appearance of EPIs can reflect the parallax structure of LFs, we applied SSIM\cite{wang2004image} that measures the similarity between two images by using the structural information on EPIs extracted from the reconstructed LFs, denoted as EPI-SSIM, from different methods to provide a quantitative evaluation towards the LF parallax structure. The average SSIM values listed in Table~\ref{timessim} show that the EPIs from the reconstructed LFs by our method have higher SSIM values, which validates that our reconstructed LFs preserve better parallax structures to some extent. %Additionally, please refer to the video demo contained in the supplementary material for observing the advantage of our method.

\subsubsection{Comparison of running time.} 
We also compared the efficiency of different methods. And the results are shown in Table~\ref{timessim}, where it can be observed that our method is faster than all methods except Yeung \textit{et al.}\cite{wing2018fast}. Note that all methods were implemented  on a desktop with Intel CPU i7-8700 @ 3.70GHz, 32GB RAM and NVIDIA GeForce RTX 2080Ti.

\subsection{Ablation Study}\label{sec:experiments_2}
\subsubsection{The number of iterative stages and SAS convolutional layers.}
In our method, the crucial step is alternately updating the $\mathbf{x}^{(t+1)}$ and $\mathbf{v}^{(t+1)}$ from the results of the $t$-th iterative stage. The number of iterative stages and SAS convolutional layers in the deep regularizer are both key factors to the reconstruction quality. Taking the task $2\rightarrow 25$ as an example, we separately carried out ablation studies on these two factors.

 \begin{table}[t]
\centering
\scriptsize
\setlength{\tabcolsep}{4.8mm}
\caption{ Running time (in second) of different reconstruction methods/average EPI-SSIM of reconstructed LFs by different methods. ``-" indicates that the method cannot work on the task}
\begin{tabular}{cccc}
\toprule
                                     & $1\rightarrow 49$             & $2\rightarrow 49$             & $4\rightarrow 49$   \\ 
\midrule
Ours (Single)                         & \textbf{80.71}/\textbf{0.935} & \textbf{80.18}/\textbf{0.981} & 80.54/\textbf{0.986} \\
Inagaki \textit{et al.}\cite{inagaki2018learning}    & 218.94/0.901                  &179.36/0.968                   & 179.36/0.973 \\
Kalantari \textit{et al.}\cite{kalantari2016learning}& -                             & 84.43/0.972                   & 168.86/0.980 \\
Yeung \textit{et al.}\cite{wing2018fast}             & -                             & -                             & \textbf{0.85}/0.974 \\
\bottomrule
\end{tabular}
\label{timessim}
\end{table}

\begin{figure}[t]
\centering
\captionsetup{skip=0pt}
\subfigure[]{\includegraphics[width=0.31\textwidth]{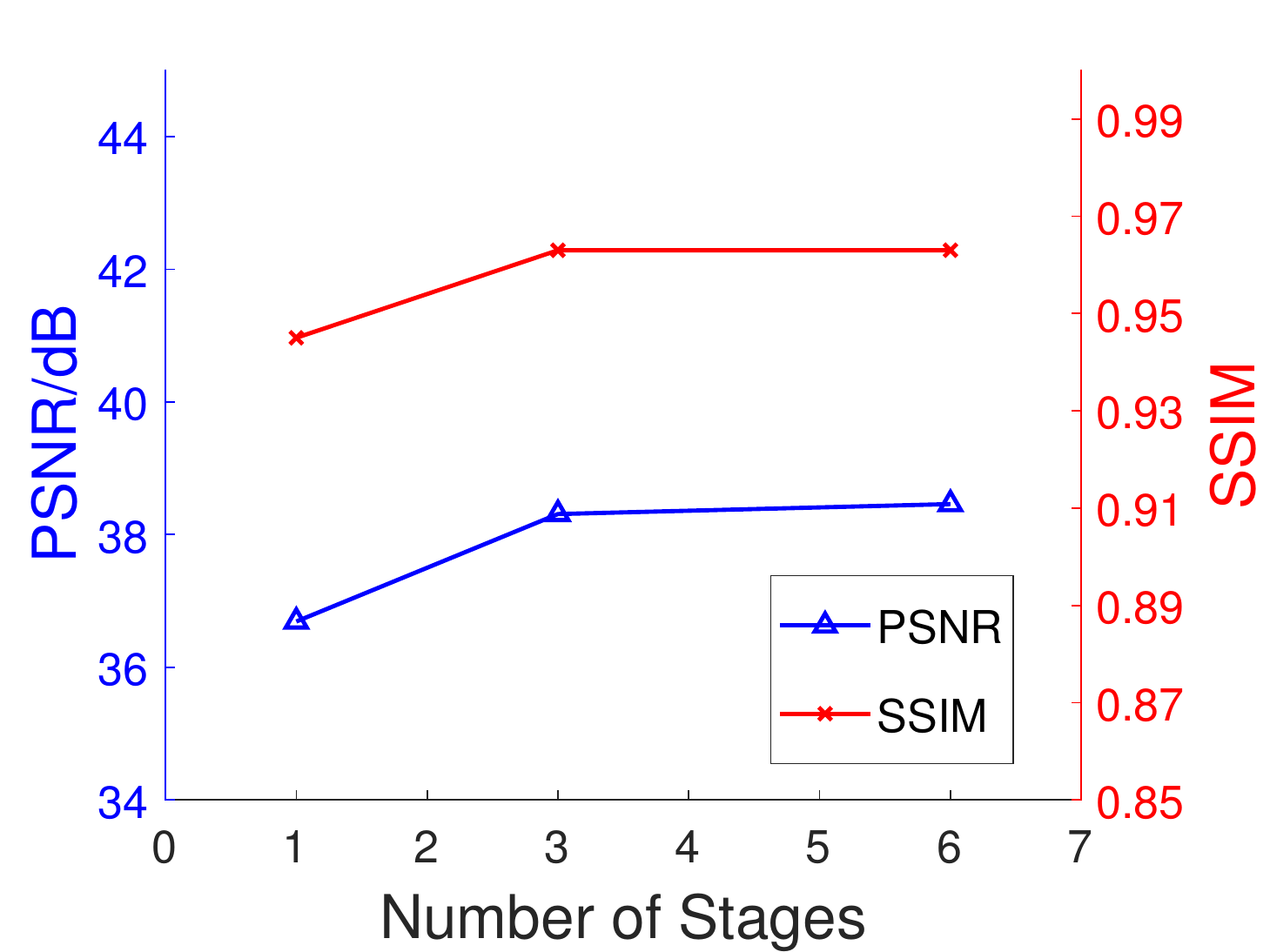}}\hfill 
\subfigure[]{\includegraphics[width=0.31\textwidth]{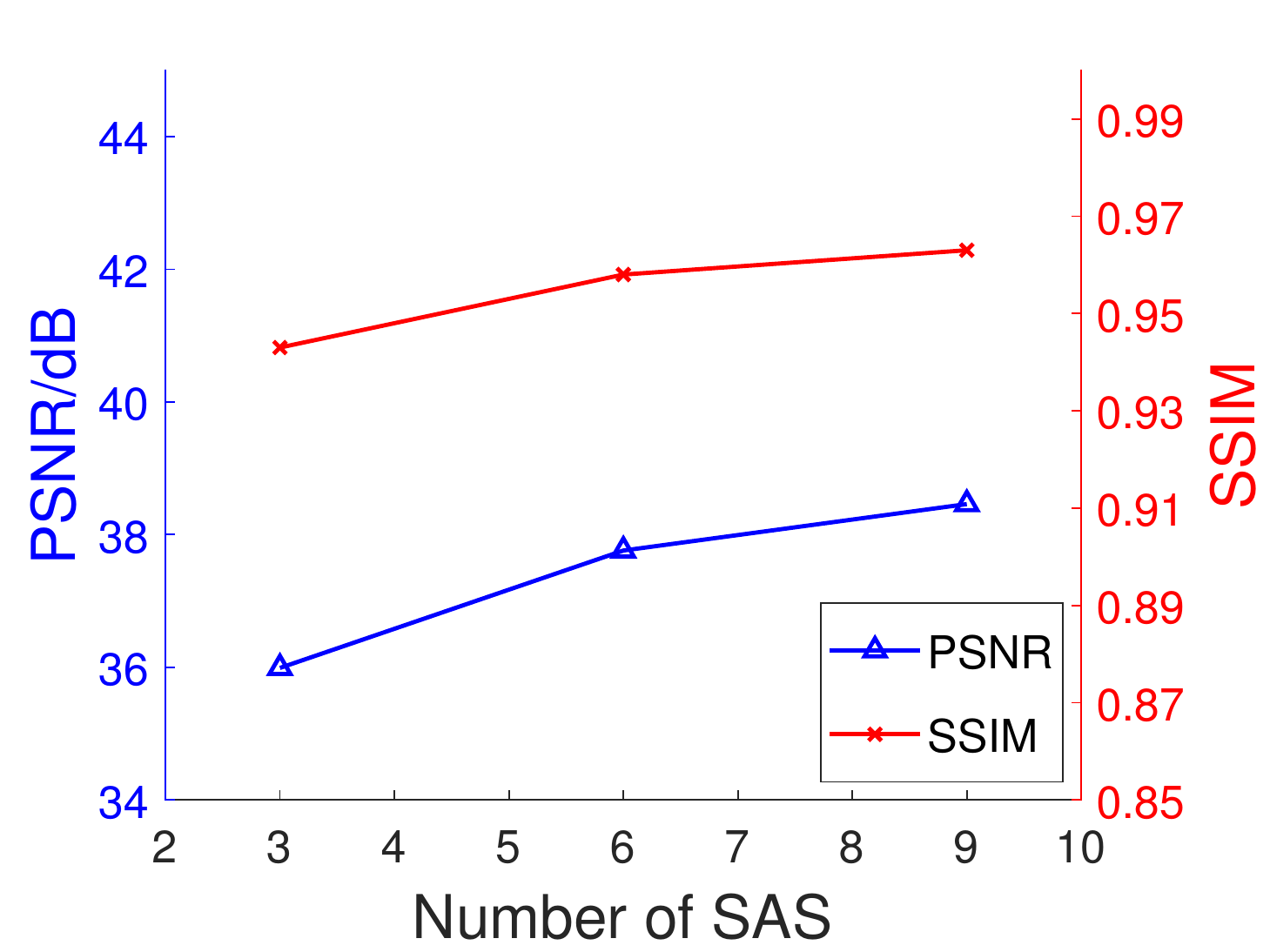}}\hfill 
\subfigure[]{\includegraphics[width=0.26\textwidth]{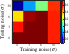}} \\
\caption{Comparisons of different numbers of iterative stages(a), different numbers of SAS convolutional layers in deep spatial-angular regularization(b), and different levels of noise(c)}
\label{abl_ps}
\end{figure}

First, three iterative-stage numbers, i.e., $1,3$ and $6$, were set, while the number of SAS was fixed to $9$. The quantitative results shown in Fig.~\ref{abl_ps}(a) indicate that with the number of iterative stages increasing, the quality of reconstructed LFs improves. Besides, the improvement is more obvious from $1$ stage to $3$ stages but very slight from $3$ stages to $6$ stages. In practice, the numbers of iterative stages and SAS convolutional layers can be optimally set according to the available computational resources. Then, we set three SAS numbers, i.e., $3,6$ and $9$, and fixed the number of iterative stages to $6$. The results are shown in Fig.~\ref{abl_ps}(b). With the number of SAS convolutional layers increasing, the quality of reconstructed LFs also improves. We hence chose $6$ stages and $9$ SAS convolutional layers to trade-off the quality of LF reconstruction and the computational costs in our method.

\subsubsection{Noisy measurements.}
Due to the small size of coded aperture and low-light conditions in practice, the measurements are used to be affected by noise. In order to evaluate the robustness of our method in real applications, we added different levels of Gaussian noise onto the measurements during training and testing. We set three standard deviations for one measurement: $\sigma=3$, $\sigma=6$, and $\sigma=30$ to control the noise level. Here we took the task $2\rightarrow 25$ on the synthetic LFs as an example. The 
PSNR value shown in Fig.~\ref{abl_ps}(c) indicate that our method with noisy inputs can preserve the comparable performance on the noise-free case when the testing noise level is lower or slightly higher than that of training. If the noise level is too high (e.g., $\sigma=30$), the performance declines rapidly.

\section{Conclusion and Future Work}\label{sec:conclusion and future work}
We proposed a novel deep learning-based LF reconstruction method from coded aperture measurements, which links the observation model of measurements and deep learning elegantly, making it more physically interpretable. To be specific, we design a deep regularization term with an efficient spatial-angular convolutional sub-network to implicitly and comprehensively explore the signal distribution. Extensive experiments over both real and synthetic datasets demonstrate that our method outperforms state-of-the-art approaches to a significant extent both quantitatively and qualitatively.
\subsubsection{Acknowledgements.} This work was supported in part by the Hong Kong RGC under Grant 9048123 (CityU 21211518), and in part by the Basic Research General Program of Shenzhen Municipality under Grant JCYJ20190808183003968.

\clearpage
\bibliographystyle{splncs04}
\bibliography{egbib}
\end{document}